\documentclass[prl,nofootinbib,twocolumn,floatfix,showpacs]{revtex4}
\usepackage{graphicx}
\usepackage{amsmath}
\usepackage{amssymb}
\bibpunct[, ]{[}{]}{,}{a}{,}{,}
\ifnum\lefthyphenmin<2\lefthyphenmin=2\fi
\ifnum\righthyphenmin<2\righthyphenmin=2\fi

\begin{document}
\title{Supercoil formation in DNA denaturation}

\author{A. Kabak\c{c}{\i}o\u{g}lu}
\affiliation{Department of Physics, Ko\c c University, Sar\i yer 34450 \. Istanbul, Turkey}
\author{E. Orlandini}
\affiliation{Dipartimento di Fisica,  CNISM  and  Sezione INFN, Universita' di
  Padova, Via Marzolo 8, 35131 Padova, Italy}
\author{D. Mukamel}
\affiliation{Department of Physics of Complex Systems, The Weizmann Institute
  of Science, Rehovot 76100, Israel}

\date{\today }

\begin{abstract}
  We generalize the Poland-Scheraga (PS) model to the case of a circular DNA,
  taking into account the twisting of the two strains around each other.
  Guided by recent single-molecule experiments on DNA strands, we assume that
  the torsional stress induced by denaturation enforces formation of
  supercoils whose writhe absorbs the linking number expelled by the loops.
  Our model predicts that, when the entropy parameter of a loop satisfies $c
  \le 2$, denaturation transition does not take place. On the other hand for
  $c>2$ a first-order denaturation transition is consistent with our model and
  may take place in the actual system, as in the case with no supercoils.
  These results are in contrast with other treatments of circular DNA melting
  where denaturation is assumed to be accompanied by an increase in twist
  rather than writhe on the bound segments.

\end{abstract}


\maketitle

Thermal denaturation of double stranded DNA~\cite{denat_ref} has been of
recent interest due to its relevance to protein synthesis, polymerase chain
reaction (PCR) and microarray technologies. The denaturation transition has
been extensively studied theoretically mainly by means of two models: $(a)$
the Poland-Scheraga (PS) model~\cite{PS} which considers the opposite bases to
be either bound with a certain energy gain or unbound (when
part of a ssDNA loop) and ignores the twisting of the strands around each
other; and $(b)$ the Peyrard-Bishop model~\cite{PB} which is a one dimensional
model in which complementary bases interact by a distance dependent
potential. Both models have been used to study, e.g., the nature of the
melting transition~\cite{PS,PB,Kafri,Fisher-Kafri2-Barbi} and the bubble
dynamics of the DNA~\cite{Barbi2,Mukamel2,Livi,Metzler}. It has recently
been shown that the PS model predicts a first-order melting transition if one
properly takes into account the self-avoidance of the
chains~\cite{Kafri,Orland1,Carlon-Causo}.

The original PS model treats the DNA as a long ladder, without considering the
twisting of the two strains around each other. It can be argued that this
feature is irrelevant for the thermodynamics of an open ended chain, since the
twisting strain can be released by the rotation of the chain ends. However,
this assumption is no longer appropriate if the chain ends are not free to
rotate or alternatively, for circular DNAs such as plasmids in bacteria. In
this case, upon heating up to the melting temperature the two strands can no
more fully depart from each other, since the chemical bonds that assemble the
sugar backbone are still intact. We assume topoisomerases and other topology
modifying agents are not present in the solution.

Then, the partition function is restricted to a sum over configurations with a
fixed {\it linking number} (the number of times one strand rotates around the
other). A well-known theorem \cite{LN_ref} states that the linking number (LN)
is the sum of {\it twist} and {\it writhe}, where twist refers to the sum of
the subsequent stacking angles along the DNA and writhe is associated with the
geometry of the DNA's center line and measures the amount of twist absorbed by
the excursions of the backbone.

Past attempts to include DNA's helicity in the PS model have considered a
twisting strain associated with the modified stacking angle of subsequent base
pairs accumulating upon loop formation.  This assumption leads to the
conclusion that the transition either changes its nature (becomes of higher
order) or disappears altogether, depending on the treatment of the
self-avoidance effects~\cite{Bruinsma_and_Rudnick} and the applied external
torsion~\cite{Garel-Orland}.  In both works, as well as here, it is assumed
that the LN accommodated by the loops is insignificant.The reason is that the
single strands may be considered as random chains and their winding angle is
rather small~\cite{Drossel}.

Recent experiments on single DNA chains, however, point to a different
possible mechanism for the absorption of LN expelled from the denaturated
loops. Experimental measurements of the torsional response of long strands
under fixed stretching force $F_s$ show that when $F_s$ is small, the chain
almost immediately undergoes a buckling transition, forming a supercoil that
absorbs the externally introduced LN~\cite{single_molecule}.  This mechanism
of harboring the LN in the modified conformation of the center line (writhe)
rather than in augmented basepair stacking angles is a familiar phenomenon
frequently observed in telephone chords. The presence of a similar phenomenon
in DNA calls for a re-examination of thermal denaturation in a setting where
the LN associated with the loops is transferred to supercoils formed by
locally relaxed DNA segments. Indeed, recent experiments point to the
presence of such a mechanism during plasmid denaturation~\cite{jpn_exp, nucl_acid}.

\begin{figure}[h]
\vspace*{0.5cm}
  \begin{center}
    \includegraphics[width=8cm]{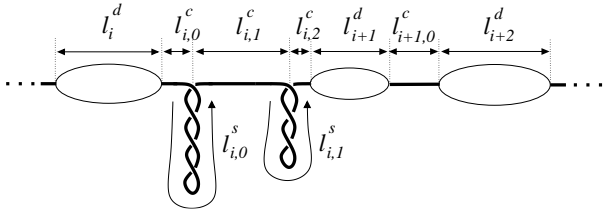}
  \end{center}
  \caption{A depiction of a microscopic configuration of the generalized Poland-Scheraga model discussed in
    the text. The superscripts (d),(c) and (s) refer to denaturation loop,
    coil and supercoil regions, respectively.}
  \label{Fig1}
\end{figure}

In Fig.~\ref{Fig1} we depict the generalized PS model we consider
for this purpose. In the model, a particular base pair may be either
unbound (in a loop), bound in a coil, or bound in a supercoil.
Unlike in the original PS model, a looped configuration is allowed
only if the rest of the chain can be rearranged to form the required
amount of supercoils for the conservation of the total LN. Let the
linking number stored in the natural twist of a relaxed DNA in
$\ell$ subsequent base-pairs be accommodated in the writhe of a
supercoil segment of total length $\ell^\prime=\delta \ell$. We will
set $\delta=1$ below, since a general treatment, although possible,
does not bring new insight. Furthermore, it is reasonable to expect
that loops form on the coils but not on the supercoils which are
relatively rigid structures. As a result, formation or expansion of
a loop on a coil is accompanied by an equal increase in the total
length of supercoils in the system.

Let $E_b<0$ be the binding energy of bound pairs in the coil or supercoil
state and $E_s >0$ be the cost of increasing the length of a supercoil by one
bp (e.g., due to the bending rigidity). For simplicity, we assume the minimum
size of a supercoil to be one base pair and one expects $E_s<-E_b$ for the
DNA. The corresponding Boltzmann factors are $\omega=\exp[-\beta E_b]$ and
$\nu = \exp[-\beta (E_b+E_s)]$, where $\beta = 1/kT$.  The Boltzmann factor
corresponding to the configuration in Fig.~\ref{Fig1} is then,
\begin{eqnarray}
&\cdots&\Omega(2l^d_i)\times w^{l^c_{i,0}}\times \nu^{l^s_{i,0}}\times
w^{l^c_{i,1}}\times \nu^{l^s_{i,1}}\times w^{l^c_{i,2}}\nonumber \\
&\times& \Omega(2l^d_{i+1})\times w^{l^c_{i+1,0}}\times
\Omega(2l^d_{i+2})\times \cdots \nonumber
\end{eqnarray}
Here $l_i^d$ represents the length, in units of base pairs, of the
loop $i$. The segment of bound pairs separating the $i$ and $i+1$
loops is composed of alternating subsegments of coiled regions of
length $l^c_{ij}$ and supercoiled regions of length $l^s_{ij}$. The
entropic contribution of a loop of length $l$ is $\Omega(2l) \equiv
A\,s^l/l^c$, where $c$ is the universal entropic parameter of a
loop, $s$ is a non-universal constant, and $A$ is a parameter which
incorporates the cooperativity parameter, i.e. the Boltzmann weight
associates with the initiation of a loop. Typically $A\simeq
10^{-4}$ and it is weakly temperature
dependent~\cite{BlosseyCarlon}.
Note that,
some inter-loop regions may accommodate several supercoils, whereas
some may have none.

Let $L_D$, $L_C$ and $L_S$ be the total length of the denaturated
regions, the coiled and the supercoiled regions, respectively,  in a
given configuration. The length of the molecule is $L=L_D+L_C+L_S$.
The canonical partition function is a sum over the contributions of
all microscopic configurations with total DNA length $L$, and with a
fixed $LN$, namely with $L_D=L_S$. In the grand-canonical ensemble,
the two constraints are relaxed  by introducing two fugacities $\mu$
and $z$, which contribute an extra weight $z^L \mu^{L_D-L_S}$ to
each configuration. The two fugacities are determined by taking the
appropriate derivative of the grand partition sum $Q$:
\begin{eqnarray}
\label{z_eqn0}
L &=& \frac{\partial\ln Q}{\partial\ln z} = \big\langle \sum_i \bigg[l^d_i +
\sum_j \big(l^c_{i,j}+l^s_{i,j}\big)\bigg] \big\rangle \\
\label{mu_eqn0} 0 &=& \frac{\partial\ln Q}{\partial\ln\mu} =
\big\langle \sum_i \bigg[ l^d_{i} - \sum_j l^s_{i,j}\bigg]
\big\rangle\ .
\end{eqnarray}
Ignoring end effects which contribute terms of order $L$ to the
partition sum, one finds that the grand partition function $Q$ can
be expressed as
\begin{eqnarray}
Q(z,\mu) &=& \tilde{V}(z,\mu) +
\tilde{V}(z,\mu)U(z\mu)\tilde{V}(z,\mu) +
\tilde{V}U\tilde{V}U\tilde{V} + \cdots \nonumber \\
&=& \tilde{V}/(1-U\tilde{V})\ ,
\label{gpf}
\end{eqnarray}
where, following Ref.~\cite{Kafri},
\begin{eqnarray}
\label{eqnU}
U(z\mu) &\equiv& \sum_{l=1}^{\infty} \Omega(2l)(z\mu)^l = A\,\Phi_c(sz\mu)\ , \\
\tilde{V}(z,\mu) &\equiv& V(z)/[1-W(z/\mu)V(z)]\ , \label{eqnvbar} \\
\mbox{with,} \ \ V(z) &\equiv& \sum_{l=1}^{\infty} (\omega z)^l = \omega z/(1-\omega z)\ ,
\nonumber\\
\label{eqnW}
W(x) &\equiv& \sum_{l=1}^{\infty} (\nu x)^l =
\frac{\nu x}{1-\nu x}\ .
\end{eqnarray}
The functions $U$, $V$, and $W$ represent the grand sums for a loop, a coil,
and a supercoil, respectively, and $\Phi_c(x)$ is the Polylog function. Note
that, the functional form of the grand sum in Eq.(\ref{gpf}) is similar to
that of the original Poland-Scheraga model, except that the ``propagator'' for
the coil regions is now dressed to accommodate an arbitrary number of
supercoils. The price paid for conserving the linking number is that its
associated fugacity $\mu$ needs to be calculated as a function of $z$ at each
temperature.  \\ \\ After some algebra Eq.(\ref{mu_eqn0}) reduces to the more
transparent relation $\frac{\partial W(z/\mu)}{\partial\mu} + \frac{\partial
U(z\mu)}{\partial\mu} = 0$.  Using Eqs.(\ref{eqnU}) and (\ref{eqnW}), we obtain
the following transcendental equation for $\mu(z)$:
\begin{eqnarray}
\label{mu_eqn}
\frac{\nu z}{(\mu-\nu z)^2}&=&  \frac{A}{\mu}\Phi_{c-1}(sz\mu)\ .
\end{eqnarray}
In order to study the nature of the denaturation transition (when it
exists) we consider the thermodynamic limit ($L\to\infty$) by
focusing on the relevant pole of $Q(z)$ at
$U(z^*\mu^*)=1/\tilde{V}(z^*/\mu^*)$, where $\mu^* = \mu(z^*)$
through Eq.(\ref{mu_eqn}). Substituting $1/\tilde{V} = 1/V - W$, we
obtain
\begin{eqnarray}
\label{thdyn_limit}
\bigg(\frac{1}{\omega z^*} - 1\bigg) - \frac{\nu z^*}{\mu(z^*)-\nu z^*} &=& A\,\Phi_c(sz^*\mu(z^*))\ .
\end{eqnarray}
The average bound pair density in coiled and supercoiled segments is given by
\begin{eqnarray}
\label{order_parameter}
\theta_c &=&-\frac{\partial \log z^*}{\partial \log \omega}\ \ \ \ \mbox{and}\
\ \ \ \ \theta_s = -\frac{\partial \log z^*}{\partial \log \nu}\ ,
\end{eqnarray}
respectively.  Therefore, a phase transition is associated with a
singularity in the temperature dependence of the solution $z^*$ of
Eqs.(\ref{mu_eqn}) and (\ref{thdyn_limit}).

Before proceeding, let us consider the simpler picture where the DNA chain is
not circular, but, nevertheless, can form supercoils.  Assuming the ends are
bound and free to rotate, we set $\mu=1$ in the grand sum, since the system is
now insensitive to the linking number. Eq.(\ref{thdyn_limit}) alone suffices
to describe the phase transition in this case.  Let the LHS and the RHS of
Eq.(\ref{thdyn_limit}) with $\mu=1$, be named $F(z)$ and $G(z)$,
respectively. Then, $F(z)$ is a smooth, monotonically decreasing function of
$z$ for $0<z<1/\nu$, with $F(0^+)=+\infty$. Similarly, $G(z)$ is a smooth,
monotonically increasing function of $z$ for $0<z\le 1/s$ (and divergent for
$z>1/s$), with $G(0)=0$. A phase transition exists if the {\it smallest} $z^*$
that satisfies Eq.(\ref{thdyn_limit}) exhibits a singularity as a function of
temperature. This is the case only if $c>1$ (so that $\Phi_c(sz)$ remains
finite as $sz\to 1$). In addition one requires $F(1/s)>G(1/s)$ at infinite
temperature where $\omega=\nu=1$. Equivalently, after substitution,
\begin{equation}
\label{upper_limit}
s - 1/(s-1) > 1 + A\zeta_c\ ,
\end{equation}
where $\zeta_c\equiv\Phi_{c}(1)$. Given a suitable set of the
phenomenological constants that satisfy these conditions, the model
exhibits a phase transition of first (second) order for $c>2$
($1<c\le2$). This mechanism is described in detail in
\cite{PS,Kafri}.

\begin{figure}[h]
\vspace*{0.5cm}
  \begin{center}
    \includegraphics[width=5cm]{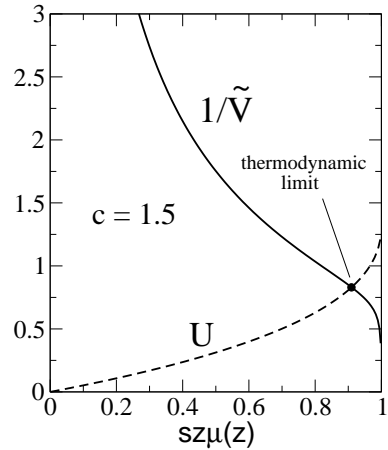}
  \end{center}
  \caption{Absence of a melting transition for $c<2$. The pole of the grand
    sum varies smoothly at all temperatures, since $\tilde{1/V}\to -\infty$ at
    $sz\mu=1$. The shown plot of $1/\tilde{V}$ is qualitatively same for all
    temperatures.}
  \label{Fig2}
\end{figure}

Having established the existence of a phase transition in the unrestricted
case and with the possibility of supercoils, let us now turn to the
circular DNA, where $\mu$ is determined by Eq.(\ref{mu_eqn}) in
order to ensure LN conservation.
For a circular DNA, two regimes emerge:

{\bf For} ${\bf 1<c\le 2}$, a solution of Eqs.(\ref{mu_eqn}) and
(\ref{thdyn_limit}) with $sz\mu=1$ does not exist, since Eq.(\ref{mu_eqn})
with $\Phi_{c-1}(1)=\infty$ dictates $\mu(z)=\nu z$, whereas in
Eq.(\ref{thdyn_limit}), $\Phi_c(1) <\infty$. In fact, $1/\tilde{V}(z,\mu)\to
-\infty$ at $sz\mu=1$ for all temperatures, ensuring a smooth variation of
$z^*$ as a function of temperature (Fig. \ref{Fig2}). Hence, the second-order
melting transition found for the unrestricted DNA is absent when the linking
number is conserved.

\begin{figure}[h]
\vspace*{0.5cm}
  \begin{center}
    \includegraphics[width=8cm]{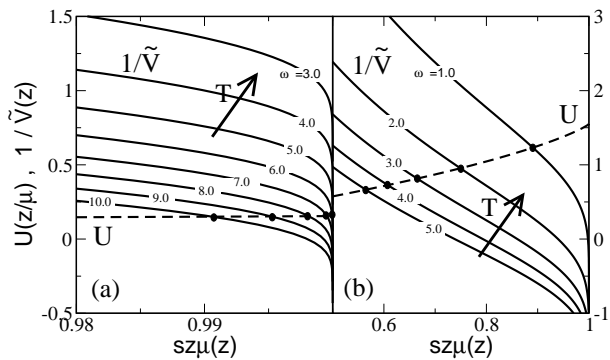}
  \end{center}
  \caption{Two melting scenarios in the model with $c=2.115$, $s=4.68$,
    $E_s=0$, and (a) $A=0.1$ with a first-order transition at
    $\omega_{crit}\simeq 4.18$ as found from Eq.(\ref{transition_temp_eq}),
    (b) $A=1.0$ with no phase transition. Full circles are the simultaneous
    solutions of Eqs.(\ref{mu_eqn},\ref{thdyn_limit}) corresponding to the
    thermodynamic limit. These parameter values which are relevant for
    self-avoiding walks on a cubic lattice~\cite{Carlon-Causo} suitably
    demonstrate the two possible scenarios, but are not intended to fit
    the experiments directly.}
  \label{Fig3}
\end{figure}

{\bf For} ${\bf c>2}$,  substituting $sz^*\mu^*=1$ in Eq.(\ref{mu_eqn}) and
Eq.(\ref{thdyn_limit}), and picking the smallest positive solution for $z^*$,
one obtains
\begin{eqnarray}
\label{star_eq}
z^* &=&  \sqrt{\frac{1}{s\nu}}\  C_-\ \ \ \mbox{and}\ \ \ \ \
\mu^* = \sqrt{\frac{\nu}{s}} \ C_+ \ \ ,
\end{eqnarray}
with $C_\pm \equiv \sqrt{1+1/(4A\zeta_{c-1})} \pm
 \sqrt{1/(4A\zeta_{c-1})}$. The critical temperature is given by
\begin{eqnarray}
\frac{\nu_{crit}^{1/2}}{\omega_{crit}} &=& \frac{C_-}{\sqrt{s}}\big[1+A\zeta_c +
C_-\sqrt{ A \zeta_{c-1}}\big]\ .
\label{transition_temp_eq}
\end{eqnarray}
Since $\nu^{1/2}/\omega \ge 1$, a phase transition exists only if $s$ is
sufficiently large. In particular, if
\begin{equation}
\label{transition_cond}
s > A\zeta_{c-1}
\end{equation}
when $A \ll 1$. The transition temperature is reduced relative to the unrestricted
case roughly by a factor $\sim \log_s(s/A\zeta_{c-1})$. With $s\approx e^{12.5}$ as used by the MELTSIM
scheme~\cite{Garel-Blake} one expects a transition to take place at a
finite temperature (although the values of $s$ and $A$ optimized for our model
may be different).  The two qualitatively different regimes separated by the
boundary in (\ref{transition_cond}) are depicted in Fig.(\ref{Fig3}).

We conclude that the circular DNA undergoes a
denaturation transition only if $c>2$.
Performing a variational analysis near $sz^*\mu^*=1$, it is
straightforward to show that the phase transition is of first order
for all values of $c>2$. Also note that, the fraction of bound pairs in coiled
and supercoiled segments, $\theta^>_c$ and $\theta^>_s$, in the high
temperature phase follow from Eqs.(\ref{order_parameter}) and
(\ref{star_eq}) as
\begin{eqnarray}
\theta^>_c &=&
0\ \ \ \ \ \mbox{and} \ \ \ \ \
\theta^>_s =
1/2\ .
\end{eqnarray}
The high temperature phase, although devoid of coiled regions, is not fully unbound. Half
of the chain is in a supercoiled state while the remaining half unbinds,
maximizing the amount of entropically favored loops.

These results contrast with the solution of the PS model without
twist~\cite{Kafri}, as well as with the earlier calculations on the
effect of locked-in twist on DNA melting
transition~\cite{Bruinsma_and_Rudnick,Garel-Orland}. The no-twist PS
model predicts a second order denaturation transition for $1<c\le
2$, whereas if supercoil formation is taken into account, this
second-order transition is absent when the LN is conserved. Earlier
attempts to incorporate twist, where the loop formation is penalized
by the overtwisting of the bound segments, found that the
first-order transition which exists for $c>2$ in the original PS
model becomes of higher order. In the proposed supercoiling
scenario, the transition for $c>2$ remains first order.

Our findings should also apply to long DNA chains, where the twist
expulsion through the ends may be hindered by kinetic effects (see, e.g.,\cite{Baiesi-Livi}).
A more general framework where both supercoiling and twisting
effects are incorporated is called for, in order to confirm the free
energetic preference for the proposed mechanism over the alternative
overtwisting scenarios.

We acknowledge helpful discussions with M. Baiesi, T. Garel, M. Hinczewski,
Y. Kafri, A. Mostafazadeh, F. Ozturk, A. Stella and C. Vanderzande. This work
was partially supported by the Scientific and Technological Research Council
of Turkey (TUBITAK) through the grant TBAG-108T553 and by the Minerva
Foundation with funding from the German Ministry for Education and
Research. We are in debt with ITAP for a school where the present
collaboration with D.M. was initiated.

\end{document}